\documentclass[a4paper,10pt]{autart}
\usepackage[T1]{fontenc}
\usepackage[utf8]{inputenc}
\usepackage[english]{babel}

\usepackage{fancybox,shadow,xcolor} 
\usepackage{setspace}
\usepackage{subfig}
\usepackage{placeins}
\usepackage{interval}
\usepackage{xcolor}
\usepackage{psfrag}
\usepackage{bm}
\usepackage{float}

\usepackage{graphicx}
\usepackage{amsmath,amssymb}
\usepackage{mathrsfs}
\usepackage{mathdots}
\usepackage{algorithmic,algorithm}

\DeclareMathOperator*{\argmax}{arg\,max}

\DeclareMathOperator*{\mce}{mce}

\DeclareMathOperator*{\suchthat}{ s.t. }

\renewcommand{\Re}{\mathbb{R}}

\usepackage{bbm}
\renewcommand{\paragraph}[1]{\smallskip\noindent\textbf{#1.} }

\newcommand{\BM}{\begin{bmatrix}}
\newcommand{\EM}{\end{bmatrix}}

\newcommand{\BBM}{\big[\begin{matrix}}
\newcommand{\EEM}{\end{matrix}\big]}

\newcommand{\bbm}{[\begin{matrix}}
\newcommand{\eem}{\end{matrix}]}

\begin{document}

\begin{frontmatter}

\title{Robustness analysis of a Maximum Correntropy framework for linear regression}
%%%%%%%
\author[AMP-ECL-UnivLyon]{Laurent Bako}
\thanks[footnoteinfo]{This paper was not presented at any IFAC 
meeting. Corresponding author L. Bako. Tel.: +33 472 186 452.}
\ead{laurent.bako@ec-lyon.fr}
\address[AMP-ECL-UnivLyon]{Laboratoire Amp\`{e}re -- Ecole Centrale de Lyon -- Universit\'{e} de Lyon, France}

%Fax .......}

\maketitle 
\setstretch{1}

\begin{abstract}
In this paper we formulate  a solution of the robust linear regression problem in a general framework of correntropy maximization.
Our formulation yields a unified class of estimators which includes the Gaussian and Laplacian kernel-based correntropy estimators as special cases. An analysis of the  robustness properties is then provided. The analysis includes a quantitative characterization of the informativity degree of the regression which is appropriate for studying the stability of the estimator. Using this tool, a sufficient condition is expressed under which  the parametric estimation error is shown to be bounded. Explicit expression of the bound is given and discussion on its numerical computation is supplied. For illustration purpose, two special cases are numerically studied. 
\end{abstract}

\end{frontmatter}

\setstretch{.97}
\section{Introduction} 
Given a set of empirical observations generated by a system along with a class of parameterized candidate models, a parameter estimator is a function which maps the available data to the parameter space associated with the model class. 
 A very desirable property for an estimator is that of robustness which characterizes a relative insensitivity of the estimator to deviations of the observed data from the assumed model. More specifically, this property is central in situations where the data are prone to non Gaussian noise or  disturbances of possibly arbitrarily large amplitude (often called outliers). The quest for robust estimators has led to the development of many estimators such as the Least Absolute Deviation (LAD) \cite{Rousseeuw05-Book,Maronna06-Book,Candes06-IT,Bako16-Automatica}, the least median of squares \cite{Rousseeuw84}, the least trimmed squares \cite{Rousseeuw05-Book}, the class of M-estimators \cite{Huber-Book-09}. Evaluating formally to what extent a given estimator is robust requires setting a quantitative measure of robustness. Incidentally such a measure can serve as comparison criterion between different robust estimators. 
Generally, the robustness is assessed in term of the maximum proportion of outliers in the total data set that the estimator can handle while remaining stable (see for example the concept of breakdown point \cite{Rousseeuw05-Book}). 
More recently the maximum correntropy \cite{Santamaria06-TSP,Liu07-TSP,Principe10-Book} has emerged as an information-theoretic estimation framework which induces some robustness properties with respect to outliers. Although maximum correntropy estimation is closely related to M-estimation, its discovery has broadened the horizon of possibilities for designing robust identification schemes. As a matter of fact, it has been successfully applied to a variety of estimation problems such as linear/nonlinear regression, filtering, face recognition in computer vision \cite{Feng15-JMLR,Chen14-SPL,He11-PAMI}.

\paragraph{Contribution}
Although the maximum correntropy based estimators have been gaining an increasing success, the formal analysis of its robustness properties is still a largely open research question. In this paper we propose such an analysis for a class of maximum correntropy based estimators applying to linear regression problems. More precisely, the contribution of the current paper is articulated around the  following three questions:  
\begin{itemize}
	\item To what extent the maximum correntropy estimation framework is robust to outliers?
	By robustness, it is meant here a certain insensitivity of the estimator to large errors of possibly arbitrarily large magnitude. To address this question, we derive parametric  estimation error bounds induced by the estimator in function of both the degree of richness of the regression data and on the fraction of  outliers. In summary, we show that if the regression data enjoy some richness properties and if the number of outliers is reasonably small, then the  parametric estimation error remains stable. Indeed the proportion of  outliers that the estimator is capable to correct depends on how rich the regressor matrix is. Moreover, the estimation error appears to be a decreasing function of the richness measure. 
	\item How does richness of the training data set influence the robustness of the estimator and how to characterize it?
	We provide an appropriate characterization of the richness in terms of the cardinality of the regressor vectors which are strongly correlated to any vector of the {regression space}. As such however, this quantitative measure of richness is not computable at an affordable price. To alleviate this difficulty the paper proposes some estimates of this measure thus allowing for the approximation of the parametric estimation error bounds.  
\item Does the maximum correntropy estimator (MCE) possess the exact recovery property?
%That is, in the scenario when the data is assumed to be contaminated only by a few arbitrary nonzero gross errors.  \\
We show that unlike the LAD estimator, the MCE  is not able to return exactly the true parameter vector once the measurement is affected by a single arbitrary nonzero error. The proof is given for the Gaussian kernel based estimator. 
%The intuitive reason for this is that ......
\end{itemize}
 We note that an analysis of robustness of the maximum correntropy has been presented recently in \cite{Chen16-TNNLS,Chen17}. However the analysis there is limited to the Gaussian kernel based correntropy and to  a single parameter estimation problem. Moreover these works do not make clear how the properties of the data contribute to the robustness of the estimator.

\paragraph{Outline} The rest of this paper is organized as follows. Section \ref{sec:pbm} presents the robust regression problem and define the class of maximum correntropy estimators whose properties are to be studied in the paper. It also introduces the general setting of the paper. The main analysis results are developed in Section \ref{sec:analysis}. In Section \ref{sec:simulations} we run numerical experiments to illustrate the richness measure and the evolution of the derived error bounds with respect to the amount of noise.   Finally,  Section \ref{sec:conclusion} contains concluding remarks concerning this work.

\paragraph{Notations}
$\Re$ is the set of real numbers; $\Re_+$ is the set of real nonnegative numbers; $\mathbb{N}$ is the set of natural integers; $\mathbb{C}$ denotes the set of complex numbers.  $N$ will denote the number of data points and $\mathbb{I}=\left\{1, \ldots, N\right\}$ the associated index set.  For any finite set $\mathcal{S}$, $\left|\mathcal{S}\right|$ refers to the cardinality of $\mathcal{S}$. However, whenever $x$ is a real (respectively complex) number, $|x|$ will refer to the absolute value (respectively modulus) of $x$.  For $x=\bbm x_1 & \cdots & x_n\eem^\top\in \Re^n$, $\left\|x\right\|_p$ will denote the $p$-norm of $x$ defined by $\left\|x\right\|_p=(|x_1|^p+\cdots+|x_n|^p)^{1/p}$, for $p\in \left\{1,2\right\}$, $\left\|x\right\|_\infty=\max_{i=1,\ldots,n}\left|x_i\right|$. The exponential of a real number $z$ will be denoted $\exp(z)$ or $e^z$ according to visual convenience; $\ln(z)$ is the natural logarithm function. For a square and positive semi-definite matrix $A$, $\lambda_{\min}(A)$ and $\lambda_{\max}(A)$ denote respectively the minimal and maximal eigenvalues of $A$.

\section{Robust regression problem}\label{sec:pbm}

\subsection{The data-generating system}
Let $\left\{x_t\right\}_{t\in \mathbb{N}}$ and $\left\{y_t\right\}_{t\in \mathbb{N}}$ be some stochastic processes taking values respectively in $\Re^n$ and $\Re$. They are assumed to be related by an equation of the form  
\begin{equation}\label{eq:system}
	y_t = x_t^\top \theta^o+v_t, 
\end{equation}
where  $\left\{v_t\right\}_{t\in \mathbb{N}}$ represent an unobserved error sequence; $\theta^o\in \Re^n$ is an unknown parameter vector. 
Eq. \eqref{eq:system} may describe a static (memoryless) system or a dynamic one. In the latter case, we will conveniently assume that the so-called regressor (or explanatory vector) $x_t$ has the following structure $x_t=\bbm u_t & u_{t-1} & \cdots & u_{t-(n-1)}\eem^\top$, i.e., \eqref{eq:system} is an FIR-type (Finite Impulse Response) system, with $u_t$ then denoting its input signal at time $t$.   

\begin{assum}\label{assum:ergodicity}
 The joint stochastic  process $\left\{(x_t,v_t)\right\}_{t\in \mathbb{N}}$ 
is independently {and} identically distributed. 
\end{assum}
{
While this assumption can hold naturally for a static system, it might not be satisfied in some practical situations. For example, if \eqref{eq:system} is a dynamic system (for instance, of FIR-type), this assumption is not satisfied\footnote{Indeed this assumption can be relaxed to an appropriate notion of stationarity and ergodicity for the joint process $\left\{(x_t,v_t)\right\}$}. But as will be seen, its only role is to highlight the correntropic origin of the estimation framework considered in this paper. 
%But as will be seen, it plays a quite minor role in this paper.  
} 
\begin{assum}
The noise sequence  $\left\{v_t\right\}$ satisfies the following: there is $\varepsilon\geq 0$ such that if we define the index sets $I_{\varepsilon}^0=\left\{t:\left|v_t\right|\leq \varepsilon\right\}$ and $I_{\varepsilon}^c=\left\{t:\left|v_t\right|> \varepsilon\right\}$, then the cardinality of $\left|I_{\varepsilon}^0\right|$ is "much larger" than that of $\left|I_{\varepsilon}^c\right|$.
\end{assum}
We will formalize latter in the paper what "much larger" can mean. 
Similarly as in \cite{Bako16-Automatica}, we can assume that $v_t$ is of the form $v_t=f_t+e_t$ where  $\left\{f_t\right\}$ is a sparse noise sequence in the sense that only a few elements of it are different from zero. However its nonzero elements are allowed to take on arbitrarily large values (called in this case, outliers). {As to $\left\{e_t\right\}$, it is assumed to  be a bounded  and dense (i.e., not necessarily sparse) noise sequence of rather moderate amplitude. }

\paragraph{Problem}
Given a finite collection $Z^N=\left\{(x_t,y_t)\right\}_{t=1}^N$ of measurements obeying the system equation \eqref{eq:system}, the robust regression problem of interest here is the one of finding a reliable estimate of the parameter vector $\theta^o$ despite the effect  of  arbitrarily large errors.

Let $\theta$ denote a  candidate parameter vector (PV) which we would like, ideally, to coincide with the true PV $\theta^o$. Given $x_t$ and $\theta$,  the  prediction we can make of  $y_t$  is  $\hat{y}_t(\theta)=x_t^\top \theta$. It is then the goal of the estimation method to select $\theta$ such that $y_t$ and $\hat{y}_t(\theta)$ are close in some sense for any $t$. Closeness will be be measured in term of the so-called maximum correntropy between the measured output $y_t$ and  the predicted value $\hat{y}_t(\theta)$. 

\subsection{Maximum correntropy estimation}
The correntropy is an information-theoretic measure of similarity between two arbitrary random variables \cite{Santamaria06-TSP,Liu07-TSP}. 
%%%%%%%%%%%%%%%%%%%%%%
%%%%%%%%%%%%%%%%%%%%%
More specifically, consider two random variables $Y$ and $\hat{Y}$  defined on the same probability space, and taking values in $\Re$. Let $\phi_\ell:\Re\times \Re\rightarrow \Re$ be a positive-definite kernel function on $\Re$ (see e.g., \cite[Chap. 2, p. 30]{Scholkopf02-Book} for a  definition).  The correntropy $V_{\phi_\ell}(Y,\hat{Y})$  between $Y$ and $\hat{Y}$ with respect to a kernel function $\phi_\ell$, is  defined by 
$$V_{\phi_\ell}(Y,\hat{Y})= \mathbb{E}_{Y,\hat{Y}}\big[\phi_\ell(Y,\hat{Y})\big], $$
where  $\mathbb{E}_{Y,\hat{Y}}\left[\cdot\right]$ refers to the expected value with respect to the joint distribution of $(Y,\hat{Y})$. In a more explicit form, we have  
\begin{equation}\label{eq:Vphi}
	V_{\phi_\ell}(Y,\hat{Y})=\int_{\Re} \int_{\Re}\phi_\ell(y,\hat{y})p_{Y,\hat{Y}}(y,\hat{y})dyd\hat{y}
\end{equation}
with $p_{Y,\hat{Y}}$ being the joint probability density function of $(Y,\hat{Y})$.  The correntropy constitutes a similarity measure between $Y$ and $\hat{Y}$ through the kernel $\phi_\ell$. 
 Although the original definition of correntropy in \cite{Santamaria06-TSP} fixes $\phi_\ell$ to be  the Gaussian kernel, it is indeed  possible to extend it to any positive definite kernel function. 

We consider in this paper  a kernel function of the form
\begin{equation}\label{eq:phi}
	\phi_\ell(y,\hat{y})=\exp(-\gamma\ell(y-\hat{y})),
\end{equation}
 where  $\gamma>0$ is a user-specified parameter and $\ell:\Re\rightarrow\Re_+$ is a function which satisfies the following properties:
\begin{enumerate}
	\item[P1.] $\ell$ is positive-definite: $\ell(a)\geq 0$ $\forall a$ and $\ell(a)=0$ if and only if $a=0$.
	\item[P2.] $\ell$ is symmetric: $\ell(-a)=\ell(a)$. %=\ell(\left|a\right|)$
	\item[P3.] $\ell$ is nondecreasing on $\Re_+$:  $\ell(a)\leq \ell(b)$ whenever $\left|a\right|\leq \left|b\right|$.
	\item[P4.] There exists $\alpha_\ell>0$ such that\\  $\ell(a-b)\geq \alpha_\ell \ell(a)-\ell(b)$ $\forall (a,b)\in \Re^2$. 
\end{enumerate}
{Property P4  can be interpreted as a relaxed version of the triangle inequality property for $\ell$. We can characterize a  family of functions $\ell$ satisfying P1-P4 as follows.   
\begin{lem}[Examples of functions obeying P1-P4]\label{lem:examples}
For any real number $p\geq 1$, the function $\ell_p:\Re\rightarrow\Re_+$ defined by $\ell_p(x)=\left|x\right|^p$ satisfies  the properties P1-P4. In particular, P4 is satisfied with $\alpha_{\ell}=1/2^{p-1}$. 
\end{lem}
\begin{pf}
That $\ell_p$ satisfies P1-P3 is an obvious fact. As to Property P4, it follows from convexity. In effect the convexity of $\ell_p$ implies that for all $(a,b)\in \Re^2$, $\left|a+b\right|^p/2^p=\ell_p((a+b)/2)\leq 1/2\ell_p(a)+1/2\ell_p(b)$. Multiplying by $2$ gives
$1/2^{p-1}\ell_p(a+b)\leq \ell_p(a)+\ell_p(b)$, which by replacing $a$ with $a-b$ can be seen to be equivalent to P4 with $\alpha_{\ell}=1/2^{p-1}$. \qed
\end{pf}
}
%%%%%
The correntropy maximization is an estimation framework where one tries to maximize the correntropy. 
In the regression problem stated above, we aim to find the parameter vector $\theta$ that maximizes\footnote{By Assumption \ref{assum:ergodicity}, $V_{\phi_\ell}(y_t,\hat{y}_t(\theta))$ is indeed constant i.e., independent of $t$. Hence  $t$ refers here to an arbitrary time index.} $V_{\phi_\ell}(y_t,\hat{y}_t(\theta))$, the correntropy between $y_t$ and $\hat{y}_{t}(\theta)$ with respect to the kernel $\phi_\ell$. In practice however the distribution\footnote{To be precise, the interest is in $p_{y_t,\hat{y}_t(\theta)}$ but this follows from $p_{x_t,y_t}$.}  $p_{x_t,y_t}$ is generally unknown so that one cannot evaluate the exact correntropy. As a consequence of this difficulty one would be content in practice with maximizing a sample estimate of the correntropy.   Assume that we are given a set  $Z^N=\left\{(x_t,y_t)\right\}_{t=1}^N$ of data points sampled independently from the joint distribution $p_{x_t,y_t}$.  Then in virtue of Assumption \ref{assum:ergodicity}, an estimate of the correntropy is given by 
\begin{equation}\label{eq:Vhat}
\begin{aligned}
		\widehat{V}_{\phi_\ell}(y_t,\hat{y}_t(\theta))& =\dfrac{1}{N}\sum_{k=1}^N\phi_\ell(y_k,\hat{y}_k(\theta))\\
		& =\dfrac{1}{N}\sum_{k=1}^N\exp\left[-\gamma\ell(y_k-x_k^\top \theta)\right]
\end{aligned}
\end{equation}
for all $t\in \mathbb{I}\triangleq \left\{1,\ldots,N\right\}$. 
Hence the maximum correntropy estimator (MCE) studied in this paper is the possibly set-valued map $\Psi_{\mce}:(\Re^{n}\times \Re)^N\rightarrow\Re^n$ which maps the data to a parameter space,  
\begin{equation}\label{eq:Psi-mce}
	\Psi_{\mce}(Z^N)=\argmax_{\theta\in \Re^n}\widehat{V}_{\phi_\ell}(y_t,\hat{y}_t(\theta)).
\end{equation}
{In the form \eqref{eq:Vhat}-\eqref{eq:Psi-mce} the MCE can be viewed as a particular instance of the prediction error estimation scheme \cite[Chap. 7]{LjungBook} with prediction error measured by $\exp\left[-\gamma\ell(y_k-x_k^\top \theta)\right]$. Also, the performance index \eqref{eq:Vhat} is reminiscent of the risk-sensitive estimation cost which is used in control, adaptive filtering and parameter estimation \cite{Boel02-TAC,Dey97-TAC}. But this latter approach, which roughly consists in the minimization of a sum of exponential of positive error terms, is not suitable for handling the effects of impulsive noise such as outliers.  
} \\
Although the focus of this paper is the analysis of the properties of the estimator \eqref{eq:Psi-mce}, let us mention in passing that the underlying optimization problem in \eqref{eq:Psi-mce} is non convex. This implies that solving \eqref{eq:Psi-mce} numerically can be challenging.   However it can be interpreted iteratively as a weighted least squares problem in the case for example where $\phi_\ell$ is taken to be the Gaussian kernel. We will get back to this in Section \ref{sec:simulations}.

\section{Robustness properties of the MCE}\label{sec:analysis}

%%%%%%%%%%%%%%%%
As discussed in the introduction, an estimator of the form \eqref{eq:Psi-mce} is intuitively thought (and empirically shown) to be endowed with some robustness properties. By this, we mean that it is able to keep behaving reasonably well when a certain fraction of the available data points are affected by  noise components  $v_t$ of possibly arbitrarily large magnitude.  The question of main interest in this paper is to characterize quantitatively up  to what extent the estimator defined in \eqref{eq:Psi-mce} can be insensitive to outliers.

\subsection{Data informativity}\label{subsec:Informativity}
As will be seen, the robustness property is inherited from both the structure of the estimator and the richness of the regression data. We are therefore interested in formalizing  as well  that richness and how it contributes to the robustness properties of the estimator.

To proceed with the analysis, let us introduce some notations. 
For convenience we  make the following assumption. 
\begin{assum}\label{eq:x-neq-0}
The regressor sequence $\left\{x_t\right\}$ satisfies: $x_t\neq 0$ for all  $t\in \mathbb{I}$. 
\end{assum}
\noindent Note that  Assumption \ref{eq:x-neq-0} is without loss  of generality. Under this assumption, let us pose 
\begin{equation}\label{eq:rx}
	r_x=\min_{t\in \mathbb{I}}\left\|x_t\right\|_2>0,
\end{equation}
with $\left\|\cdot\right\|_2$ denoting Euclidean norm. 
Upon dividing the system equation \eqref{eq:system} by $\left\|x_t\right\|_2$, we can even assume that $r_x=1$. 
Let $\alpha \in \interval[]{0}{1}$ be a real number. For any  $\eta\in \Re^n$, define the index set 
\begin{equation}\label{eq:I-alpha-X}
	\mathscr{I}_\alpha(X,\eta)=\left\{t\in \mathbb{I}: |x_t^\top \eta|\geq \alpha \left\|x_t\right\|_2 \left\|\eta\right\|_2\right\}
\end{equation}
with $X = \bbm x_1 & \cdots & x_N\eem\in \Re^{n\times N}$ a matrix formed with all the regressors.
Finally, let $\rho_{\alpha}(X)$ be the ratio between  the minimum cardinality that $\mathscr{I}_\alpha(X,\eta)$ can attain over all possible values of $\eta$, and the number $N$ of columns in $X$, i.e., 
\begin{equation}\label{eq:Rho}
	\rho_{\alpha}(X)= \inf_{\eta\in \Re^n}\dfrac{1}{N}\left|\mathscr{I}_\alpha(X,\eta)\right|. 
\end{equation}
The number $\rho_\alpha(X)$ measures somehow the richness (or informativity/genericity) of the regression data. Intuitively,  $\rho_\alpha(X)$ reflects a dense spanning of all directions of the vector space $\Re^n$ by the  vectors $\left\{x_t\right\}$.  For a given  $\alpha>0$, it is desired that  $\rho_\alpha(X)$ be as large as possible.   We will refer to it as the correlation measure of the  matrix $X$ at the level $\alpha$. 

It appears intuitively  that $\rho_\alpha(X)$ is a decreasing function of $\alpha$. Clearly, we get $\rho_\alpha(X)=0$ for $\alpha=1$ for finite $N$ while $\rho_\alpha(X)=1$  for $\alpha=0$. 
For a given matrix, it would be interesting to be able to evaluate numerically the quantitative measure $\rho_\alpha(X)$ of richness. Indeed, this value will be required for numerical assessment of the error bound to be derived in Section \ref{subsec:Main-Result}.   However computing exactly the value of $\rho_\alpha(X)$ is a hard combinatorial problem. 

\noindent We therefore discuss how to reach estimates of $\rho_\alpha(X)$ at an affordable cost. 
To this end, let $\tilde{X}=\bbm \tilde{x}_1 & \cdots & \tilde{x}_N \eem\in \Re^{n\times N}$ be the matrix obtained from $X$ by normalizing its  columns to unit $2$-norm, i.e., $\tilde{x}_t=x_t/\left\|x_t\right\|_2$ for all $t$. Then introduce the number  
\begin{equation}\label{eq:sigma}
	\sigma(X) = \min_{\eta\in \Re^n}\left\{\big\|\tilde{X}^\top \eta\big\|_{\infty}\: \suchthat \left\|\eta\right\|_2=1\right\} 
\end{equation}
which is solely a function of the matrix $X$, hence the notation. Note that the so-defined $\sigma(X)$ lies necessarily in the real interval $\interval[]{0}{1}$. Moreover,  it can be usefully observed that $\sigma(X)\geq \lambda_{\min}^{1/2}(\tilde{X}\tilde{X}^\top)/\sqrt{N}$, with $\lambda_{\min}^{1/2}(\cdot)$ referring to the square root of the minimum eigenvalue. 
Now for any $t\in \mathbb{I}$ consider the following index set 
\begin{equation}\label{eq:Jta}
	J_{t,\alpha}(X)=\left\{k\in \mathbb{I}: |\tilde{x}_k^\top \tilde{x}_t|\geq \sqrt{1-\delta^2} \right\}
\end{equation}
where $\delta= \sqrt{1-\alpha^2}-\sqrt{1-\sigma(X)^2}$. 
It is assumed in the definition \eqref{eq:Jta} that $\sigma(X)\geq \alpha$ so that $\delta\geq 0$. 
{For a given $t$, $J_{t,\alpha}(X)$ collects the indices of the regressors which are the most correlated to $\tilde{x}_t$ in the sense that the cosine of the angle they form with $\tilde{x}_t$ is larger than $\sqrt{1-\delta^2}$.} 
Finally, let 
\begin{equation}
	v_{\alpha}(X)=\dfrac{1}{N}\inf_{t\in \mathbb{I}} \left|J_{t,\alpha}(X)\right|
\end{equation}
be the ratio between the minimum cardinality of the finite set $J_{t,\alpha}(X)$ over all $t$ living in $\mathbb{I}$ and the number $N$ of columns in $X$.  
Then we can estimate $\rho_\alpha(X)$ as follows. 
%%%%%%%%%%%%%%%%%%%%%%%%
\begin{prop}
\label{prop:Estimates-Rho}
Let $X\in \Re^{n\times N}$ be a real matrix. Then, for all $\alpha\in \interval[open left]{0}{1}$ with $\alpha\leq \sigma(X)$,  
\begin{equation}\label{eq:Estimates-Rho}
	v_{\alpha}(X)\leq \rho_\alpha(X)\leq \min\Big(1,\frac{\lambda_{\min}(\tilde{X}\tilde{X}^\top)}{N\alpha^2}\Big)
\end{equation}
with $\lambda_{\min}(\tilde{X}\tilde{X}^\top)$ denoting the minimum eigenvalue of the matrix $\tilde{X}\tilde{X}^\top$.
\end{prop}
 %%%%
The proof of this proposition uses the following lemma. 
\begin{lem}\cite[Thm 5.14]{Zhang99-Book}\label{lem:Zhang}
Let $x,y,z \in \mathbb{C}^n$ be such that $x^*x=y^*y=z^*z=1$ with $x^*$ denoting the conjugate transpose of $x$. Then
$$\sqrt{1-|x^*y|^2}\leq \sqrt{1-|x^*z|^2}+\sqrt{1-|z^*y|^2}.  $$
Equality holds if and only if there exists $\beta\in \mathbb{C}$ such that either $z=\beta x$ or $z=\beta y$. 
\end{lem}
\vspace{-.2cm}
\begin{pf}[Proof of Proposition \ref{prop:Estimates-Rho}]
The upper bound is immediate. To see this, let $\eta_0\neq 0$ be the eigenvector associated with the smallest eigenvalue of $\tilde{X}\tilde{X}^\top$. 
Then 
$$\begin{aligned}
	\lambda_{\min}(\tilde{X}\tilde{X}^\top)\left\|\eta_0\right\|_2^2 %& = \eta_0^\top(XX^\top)\eta_0\\
	&=\sum_{t=1}^N(\tilde{x}_t^\top \eta_0)^2\\
	&\geq \sum_{t\in \mathscr{I}_\alpha(X,\eta_0)}(\tilde{x}_t^\top \eta_0)^2\\
	&\geq  \left|\mathscr{I}_\alpha(X,\eta_0)\right|\alpha^2  \left\|\eta_0\right\|_2^2
\end{aligned}. $$
It follows that $\rho_{\alpha}(X)=1/N\inf_{\eta\in \Re^n}\left|\mathscr{I}_{\alpha}(X,\eta)\right|\leq 1/N\left|\mathscr{I}_{\alpha}(X,\eta_0)\right|\leq \lambda_{\min}(\tilde{X}\tilde{X}^\top)/(N\alpha^2)$. The upper inequality in \eqref{eq:Estimates-Rho} follows by additionally taking into consideration the obvious fact that $\rho_\alpha(X)\leq 1$. 

We now prove the inequality $v_{\alpha}(X)\leq \rho_\alpha(X)$. To begin with, note from \eqref{eq:sigma} that  for any $\eta\in \Re^n$ satisfying $\left\|\eta\right\|_2=1$, there exists  $t(\eta)\in \mathbb{I}$ such that  $|\eta^\top\tilde{x}_{t(\eta)}|\geq \sigma(X)$. Consider an index $k\in \mathbb{I}$,  such that $|\tilde{x}_k^\top \tilde{x}_{t(\eta)}|\geq h$ for some $h\in \interval[]{0}{1}$. Then observe that $|\tilde{x}_k^\top \eta|\geq \alpha$ is equivalent to  $$\sqrt{1-(\tilde{x}_k^\top \eta)^2}\leq \sqrt{1-\alpha^2}.$$ 
 On the other hand, by applying Lemma \ref{lem:Zhang}, we can write
$$
\begin{aligned}
	\sqrt{1-(\tilde{x}_k^\top \eta)^2} &\leq \sqrt{1-(\eta^\top\tilde{x}_{t(\eta)})^2}+\sqrt{1-(\tilde{x}_k^\top \tilde{x}_{t(\eta)})^2}\\
	& \leq \sqrt{1-\sigma(X)^2}+\sqrt{1-h^2}. 
\end{aligned}
$$
It follows that  for $|\tilde{x}_k^\top \eta|\geq \alpha$ to hold, it is sufficient that 
$$\sqrt{1-\sigma(X)^2}+\sqrt{1-h^2}\leq \sqrt{1-\alpha^2}, $$
which in turn is equivalent to $h\geq \sqrt{1-\delta^2}$ with $\delta= \sqrt{1-\alpha^2}-\sqrt{1-\sigma(X)^2}\in \interval[]{0}{1}$ by the assumption that $\sigma(X)\geq \alpha$. 
Hence, for $|\tilde{x}_k^\top \eta|$ to be greater than or equal to $\alpha$, it is enough that $|\tilde{x}_k^\top \tilde{x}_{t(\eta)}|\geq \sqrt{1-\delta^2}$.  
This means that for a given  $t$,  $k$ being in the index set $J_{t(\eta),\alpha}(X)$ defined in \eqref{eq:Jta} is a sufficient condition for $|\tilde{x}_k^\top \eta|\geq \alpha$ for all $\eta\in \Re^n$ such that $\left\|\eta\right\|_2=1$. Therefore $J_{t(\eta),\alpha}(X)\subset \mathscr{I}_\alpha(X,\eta)$ hence implying that $\left|J_{t(\eta),\alpha}(X)\right|\leq \left|\mathscr{I}_\alpha(X,\eta)\right|$. Taking now the infimum produces $v_{\alpha}(X)\leq \inf_{\eta\in \Re^n}\dfrac{1}{N}\left|J_{t(\eta),\alpha}(X)\right|\leq \inf_{\eta\in \Re^n}\dfrac{1}{N}\left|\mathscr{I}_\alpha(X,\eta)\right|=\rho_\alpha(X)$. \qed
\end{pf}
\vspace{-.2cm}
%%%
\noindent The key benefit of Proposition \ref{prop:Estimates-Rho} is that it provides a method for estimating the measure $\rho_\alpha(X)$ defined in \eqref{eq:Rho} at an affordable cost. Note however that while the upper bound in \eqref{eq:Estimates-Rho} can be computed easily, obtaining the lower bound $v_{\alpha}(X)$ is still challenging. The reason is that this bound involves the number $\sigma(X)$ in \eqref{eq:sigma} whose numerical evaluation requires solving a nonconvex optimization problem. Nevertheless, it can be approximated through some heuristics, e.g. by solving a sequence of  linear programs.   
%%%
\vspace{-.2cm}
{
\begin{rem}
In comparison to the classical concept of persistence of excitation (PE) in system identification, the richness property requiring that $\rho_{\alpha}(X)$ be large is a  stronger property. In finite time, the quantitative persistence of excitation (called specifically sufficiency of excitation in this case)  asks for the condition number
$\lambda_{\max}^{1/2}(XX^\top)/\lambda_{\min}^{1/2}(XX^\top) $
of $XX^\top$ to be as close to $1$ as possible.
The PE condition appears to be a global property of the matrix $X$ while the richness condition introduced here is a somewhat local property as it is basically counting the number of vectors $x_t$ pointing  in any direction of the regression space. 
\end{rem}
}
\vspace{-.2cm}
\subsection{Main results}\label{subsec:Main-Result}
\noindent Equipped with the measure of informativity introduced above, we can now state the main result of this paper, which stands as follows. 
\vspace{-.2cm}
\begin{thm}\label{thm:main-theorem} 
Let $I_\varepsilon^0=\left\{t\in \mathbb{I}: \left|v_t\right|\leq \varepsilon\right\}$ and $I_\varepsilon^c=\left\{t\in \mathbb{I}: \left|v_t\right|> \varepsilon\right\}$ with $\left\{v_t\right\}$ denoting the noise sequence in \eqref{eq:system}.
Let $\ell$ be a function obeying P1-P4. 
Assume that the following condition is satisfied for some  $\alpha \in \interval[open left]{0}{1}$, 
\begin{equation}\label{eq:Assump-INEQ}
	\dfrac{1}{1+e^{-\gamma\ell(\varepsilon)}}\rho_\alpha(X)+e^{-\gamma\ell(\varepsilon)}\dfrac{|I_\varepsilon^0|}{N} >1. 
\end{equation}
Then for any $\theta^\star\in \Psi_{\mce}\left(Z^N\right)$ with $Z^N$ being generated by system \eqref{eq:system}, 
%\argmax_{\theta\in \Re^n}\widehat{V}_{\phi_\ell}(y_t,\hat{y}_t(\theta))$, 
it holds that
\begin{equation}
\ell(\alpha r_x \left\|\theta^\star-\theta^o\right\|_2)\leq \dfrac{1}{\gamma\alpha_\ell}\ln(1/\mu_\ell),  
\end{equation}
where %$r_x=\min_{t\in \mathbb{I}}\left\|x_t\right\|_2>0$ and 
\begin{equation}\label{eq:mul}
\begin{aligned}
	\mu_\ell = \dfrac{1+e^{-\gamma \ell(\varepsilon)}}{\frac{|I_\varepsilon^0|}{N}+\rho_\alpha(X)-1 }&\left[\dfrac{1}{1+e^{-\gamma\ell(\varepsilon)}}\rho_\alpha(X)\right. \\  & \qquad \left.+e^{-\gamma\ell(\varepsilon)}\frac{|I_\varepsilon^0|}{N}-1\right]
\end{aligned}
\end{equation}
If in addition, $\ell$ is strictly increasing on $\Re_+$, then
\begin{equation}\label{eq:Error-Bound}
	\left\|\theta^\star-\theta^o\right\|_2\leq \dfrac{1}{\alpha r_x} \ell^{-1}\left(\dfrac{1}{\gamma\alpha_\ell}\ln(1/\mu_\ell) \right).  
\end{equation}
\end{thm}
%%%%%%%%%%%%%%%%%%%%%%%%%%%%%%%%%%%%%%%%
\vspace{-1cm}
\begin{pf}
Let 
$$\theta^\star\in \Psi_{\mce}\left(Z^N\right)=\argmax_{\theta\in \Re^n}\widehat{V}_{\phi_\ell}(y_t,\hat{y}_t(\theta)).$$ 
Then for any $\theta\in \Re^n$, it holds that
$$\sum_{t=1}^N\exp\big[-\gamma\ell(y_t-x_t^\top\theta)\big]\leq \sum_{t=1}^N\exp\big[-\gamma\ell(y_t-x_t^\top\theta^\star)\big]$$
Taking in particular $\theta=\theta^o$ and invoking the system equation \eqref{eq:system}, it follows that 
$$\begin{aligned}
	\sum_{t\in I_\varepsilon^0}\exp\big[-\gamma\ell(v_t)\big]+&\sum_{t\in I_\varepsilon^c}\exp\big[-\gamma\ell(v_t)\big]\\
	&\leq \sum_{t\in I_\varepsilon^0} \exp\big[-\gamma\ell(v_t-x_t^\top\eta^\star)\big] \\
	 & \quad \quad + \sum_{t\in I_\varepsilon^c} \exp\big[-\gamma\ell(v_t-x_t^\top\eta^\star)\big] %\\
\end{aligned}
$$
where we have posed $\eta^\star = \theta^\star-\theta^o$. 
This implies that 
$$
\sum_{t\in I_\varepsilon^0}\exp\big[-\gamma\ell(v_t)\big]\leq \sum_{t\in I_\varepsilon^0} \exp\big[-\gamma\ell(v_t-x_t^\top\eta^\star)\big] + \left|I_\varepsilon^c\right|
$$
With  $\left|v_t\right|\leq \varepsilon$ for any $t\in I_\varepsilon^0$, we have $-\gamma\ell(v_t)\geq -\gamma\ell(\varepsilon)$ by the symmetry and nondecreasing properties of $\ell$. As a consequence, $\sum_{t\in I_\varepsilon^0}\exp\big[-\gamma\ell(v_t)\big]\geq \sum_{t\in I_\varepsilon^0}\exp\big[-\gamma\ell(\varepsilon)\big]$. On the other hand, by the fourth property of the function $\ell$, $\ell(v_t-x_t^\top\eta^\star)\geq \alpha_{\ell}\ell(x_t^\top\eta^\star)-\ell(v_t)\geq \alpha_\ell \ell(x_t^\top\eta^\star)-\ell(\varepsilon)$ hence implying that $-\gamma\ell(v_t-x_t^\top \eta^\star)\leq -\gamma\alpha_\ell\ell(x_t^\top \eta^\star)+\gamma\ell(\varepsilon)$. Combining these observations allows us to write 
{
\begin{equation*}
	\begin{aligned}
		& \sum_{t\in I_\varepsilon^0}\exp\big[-\gamma\ell(\varepsilon)\big]-\left|I_\varepsilon^c\right|\\
		& \: \quad \leq  \exp\big[\gamma \ell(\varepsilon)\big]\sum_{t\in I_\varepsilon^0} \exp\big[-\gamma \alpha_\ell\ell(x_t^\top\eta^\star)\big]\\
		& \: \quad = \exp\big[\gamma \ell(\varepsilon)\big]\sum_{t\in I_\varepsilon^0\cap \mathscr{I}_\alpha(X,\eta^\star)} \exp\big[-\gamma \alpha_\ell\ell(x_t^\top\eta^\star)\big]  \\ 
		&\: \qquad \: \:  +\exp\big[\gamma \ell(\varepsilon)\big]\sum_{t\in I_\varepsilon^0\cap \mathscr{I}^c_\alpha(X,\eta^\star)} \exp\big[-\gamma \alpha_\ell\ell(x_t^\top\eta^\star)\big]
	\end{aligned}
\end{equation*}
}
In the last equality we have partitioned the set $I_\varepsilon^0$ into $I_\varepsilon^0\cap \mathscr{I}_\alpha(X,\eta^\star)$ and $I_\varepsilon^0\cap \mathscr{I}^c_\alpha(X,\eta^\star)$ with $\mathscr{I}^c_\alpha(X,\eta^\star)$ being the complement of $\mathscr{I}_\alpha(X,\eta^\star)$ in $\mathbb{I}$. Note from \eqref{eq:I-alpha-X} that for all $t\in I_\varepsilon^0\cap \mathscr{I}_\alpha(X,\eta^\star)$, $\ell(x_t^\top\eta^\star)\geq \ell(\alpha r_x \left\|\eta^\star\right\|_2)$ so that $-\gamma\alpha_\ell \ell(x_t^\top\eta^\star)\leq -\gamma\alpha_\ell \ell(\alpha r_x \left\|\eta^\star\right\|_2) $. Plugging these observations into the above inequality yields 
$$
\begin{aligned}
	& \exp\big[-\gamma \ell(\varepsilon) \big]\Big\{\left|I_\varepsilon^0\right|\exp\big[-\gamma\ell(\varepsilon)\big]-\left|I_\varepsilon^c\right|\Big\}\\
	&\qquad \qquad \leq  \left|I_\varepsilon^0\cap \mathscr{I}_\alpha(X,\eta^\star)\right|\exp\big[-\gamma\alpha_\ell \ell(\alpha r_x \left\|\eta^\star\right\|_2)\big] \\
	&\qquad \qquad \qquad \qquad \qquad \qquad \qquad +\left|I_\varepsilon^0\cap \mathscr{I}^c_\alpha(X,\eta^\star)\right|. 
\end{aligned}
$$
By observing that $\left|I_\varepsilon^0\cap \mathscr{I}^c_\alpha(X,\eta^\star)\right| =|I_\varepsilon^0|-\left|I_\varepsilon^0\cap \mathscr{I}^0_\alpha(X,\eta^\star)\right|$, we can rearrange the above inequality in the form 
\begin{equation}\label{eq:INEQ}
	\begin{aligned}
		&e^{-\gamma \ell(\varepsilon)}\Big(\left|I_\varepsilon^0\right|e^{-\gamma\ell(\varepsilon)}+\left|I_\varepsilon^0\right|-N\Big)
		-\left|I_\varepsilon^0\right|\\
		&\leq  \left|I_\varepsilon^0\cap \mathscr{I}_\alpha(X,\eta^\star)\right|\Big\{\exp\big[-\gamma\alpha_\ell \ell(\alpha r_x \left\|\eta^\star\right\|_2)\big]-1\Big\}
	\end{aligned}
\end{equation}
Now by exploiting  the definition of $\rho_\alpha(X)$, we can observe  that
$$
\begin{aligned}
	|I_\varepsilon^0\cap \mathscr{I}_\alpha(X,\eta^\star)|&= |I_\varepsilon^0|+|\mathscr{I}_\alpha(X,\eta^\star)|-|I_\varepsilon^0 \cup \mathscr{I}_\alpha(X,\eta^\star)| \\
	&\geq |I_\varepsilon^0|+N\rho_\alpha(X)-N. 
\end{aligned}
$$
Moreover since  	
$$N\rho_\alpha(X)+|I_\varepsilon^0|\geq \dfrac{1}{1+e^{-\gamma\ell(\varepsilon)}}N\rho_\alpha(X)+e^{-\gamma\ell(\varepsilon)}|I_\varepsilon^0|, $$ 
the assumption \eqref{eq:Assump-INEQ} guarantees that $|I_\varepsilon^0|+N\rho_\alpha(X)-N>0$.  
Therefore since the term on the right hand side of \eqref{eq:INEQ} is negative, it holds that
$$
\begin{aligned}
&	e^{-\gamma \ell(\varepsilon)}\Big\{\left|I_\varepsilon^0\right|e^{-\gamma\ell(\varepsilon)}+\left|I_\varepsilon^0\right|-N\Big\}
	-\left|I_\varepsilon^0\right|\\
	&\leq  \left(|I_\varepsilon^0|+N\rho_\alpha(X)-N\right)\Big\{\exp\big[-\gamma\alpha_\ell \ell(\alpha r_x \left\|\eta^\star\right\|_2)\big]-1\Big\}
\end{aligned}
$$
Then direct algebraic calculations lead to
$$\mu_\ell\leq  
\exp\big[-\gamma\alpha_\ell \ell(\alpha r_x \left\|\eta^\star\right\|_2)\big]\leq 1
$$
where $\mu_\ell$ is defined as in \eqref{eq:mul}.
Indeed, in virtue of the assumption \eqref{eq:Assump-INEQ},  $\mu_\ell$ is positive.  Hence we have
$$\ell(\alpha r_x \left\|\eta^\star\right\|_2)\leq \dfrac{1}{\gamma\alpha_\ell}\ln(1/\mu_\ell).  $$
Of course, if $\ell$ is monotonically increasing on $\Re_+$ then it is invertible and the error bound in \eqref{eq:Error-Bound} follows.
\qed 
\end{pf}
%%%%%%%%%%%%%%%%%%%%%%%%%%%%%%%%%%%%%%%%%%%

%%%%%%%%%%%%%%%%%%%%%%%%%%%%%%%%%%%%%%%%%%%%%%%%%%%%%%%%%%%%%%%%%%%%%%%%%%%%%%%%%%%%
A few comments follow from this result. 
	A key assumption of the theorem is condition \eqref{eq:Assump-INEQ}. What it requires is  on the one hand,  that the proportion  of outliers be somehow small and  on the other hand, that the regression data $X$ be rich in the sense that $\rho_\alpha(X)$ be large enough for a given nonzero $\alpha \in \interval[open left]{0}{1}$. An important teaching of this condition is that the richer the data matrix $X$, the larger the number of outliers that can be corrected by the estimator. We can interpret \eqref{eq:Assump-INEQ} as a sufficient condition for the stability  of the estimator since it guarantees a bounded estimation error. 
	%\item 
	
	A second comment concerns the amplitude of the error bound given in \eqref{eq:Error-Bound}. For the purpose of making this bound small, we need the constant $\mu_\ell$ to be close to one. Again we see that this is favored by a small number of outliers and a rich data set.  An interesting special case is when $\ell(\varepsilon)=0$, which  occurs when the data are only affected by some outliers ($\varepsilon=0$) and no dense noise. In this case the number  $\mu_\ell$ defined in \eqref{eq:mul} reduces to 
$$
\mu_\ell =1- \dfrac{1-\frac{|I_\varepsilon^0|}{N}}{\frac{|I_\varepsilon^0|}{N}+\rho_\alpha(X)-1 }
$$	
which tend to suggest, since $\mu_\ell\neq 1$, that no exact recovery might  be achieved once the data are affected by a single outlier unless we consider in \eqref{eq:Error-Bound} the limit case when $\gamma\rightarrow +\infty$. {A similar observation was made in \cite{Chen17} in a comparable context. We will  prove below that the MCE does not possess the exact recovery property, at least in the case when $\ell(a)=a^2$.}  In contrast, a robust estimator such as the LAD estimator (see, e.g., \cite{Bako17-TAC,Bako16-Automatica}) is able to achieve exact recovery under a relatively significant proportion of nonzero errors. 
%\end{itemize}
%%%%%%%%%
{
\begin{prop}
 Let Assumption \ref{eq:x-neq-0} hold and assume that for any $t\in I_\varepsilon^0$, $v_t=0$. Take  the function $\ell$ in \eqref{eq:Vhat} to be the square function, $\ell(a)=a^2$. 
For all $\varepsilon> 0$, if $\left|I_\varepsilon^c\right|\geq 1$ then there exists a sequence $\left\{v_t\right\}$ such that 
$\theta^o\notin  \Psi_{\mce}(Z^N)$ when $Z^N$ is generated from \eqref{eq:system} under the action of $\left\{v_t\right\}$. 
\end{prop}
%%%%%%%%%%%
\begin{pf}
We start by observing that with $\ell$ being the square function, the cost $\hat{V}_{\phi_\ell}(y_t,\hat{y}_t(\theta))$ is differentiable.  Therefore, a necessary condition for  $\theta$ to be in $\Psi_{\mce}(Z^N)$ is that  $\nabla \hat{V}_{\phi_\ell}(y_t,\hat{y}_t(\theta))=0$, where $\nabla$ refers to the gradient. This, by using the system equation \eqref{eq:system} and exploiting the assumption that $v_t=0$ for $t\in I_{\varepsilon}^0$,  can be translated into
$$\left(\sum_{t=1}^Nw_t(\theta)x_tx_t^\top\right)\left(\theta-\theta^o\right)=\sum_{t\in I_{\varepsilon}^c} w_t(\theta)v_tx_t,$$
where $w_t(\theta)=\exp\big(-\gamma\ell(v_t-x_t^\top(\theta-\theta^o))\big)$. % with $\eta=\theta-\theta^o$. 
Note that the matrix on the left hand side of the equation above is finite regardless of the value of $\theta$. Hence, in the event that $\theta^o\in  \Psi_{\mce}(Z^N)$, we would have 
$$\sum_{t\in I_{\varepsilon}^c} w_t(\theta^o)v_tx_t=\sum_{t\in I_{\varepsilon}^c} \lambda_t x_t= 0$$
with $\lambda_t=\exp(-\gamma\ell(v_t))v_t$. Clearly, it is possible to find a nonzero sequence $\left\{v_t\right\}$ which does not meet this condition. Hence $\theta^o$ cannot be  in  $\Psi_{\mce}(Z^N)$ for an arbitrary $\left\{v_t:t\in I_{\varepsilon}^c\right\}$ no matter how small the cardinality of $I_\varepsilon^c$ is. \qed 
\end{pf}
}
%%%%%%%%%%%%%%%%%%%%
\noindent We now discuss some special instances of Theorem \ref{thm:main-theorem} corresponding to two kernels which are frequently used for estimation. For convenience of the  discussion, let us introduce the following notation. Let $\mu:\Re_+^3\rightarrow \Re_+$
$$
\mu(z,\varepsilon,\alpha)=\dfrac{1+e^{-z}}{\frac{|I_\varepsilon^0|}{N}+\rho_\alpha(X)-1 }\Big[\dfrac{1}{1+e^{-z}}\rho_\alpha(X)+e^{-z}\dfrac{|I_\varepsilon^0|}{N}-1\Big]
$$
whenever $e^{-z}\frac{|I_\varepsilon^0|}{N}+\frac{1}{1+e^{-z}}\rho_\alpha(X)-1>0$ and $\mu(z,\varepsilon,\alpha)=0$ otherwise.
%%%%%%
\begin{rem}
The bound \eqref{eq:Error-Bound} allows for some degree of freedom in the choice of the parameter $\alpha$. 
For a given function $\ell$ assumed to be invertible on $\Re_+$, a better bound can, in principle, be obtained as  
$$\min_{\alpha\in \interval[open left]{0}{1}} \dfrac{1}{\alpha r_x} \ell^{-1}\left(\dfrac{1}{\gamma\alpha_\ell}\ln\Big(\dfrac{1}{\mu(\gamma \ell(\varepsilon),\varepsilon,\alpha)}\Big) \right)$$
subject to $\alpha\leq \sigma(X)$ and condition \eqref{eq:Assump-INEQ}. Although such a minimum might not be easy to compute exactly, one can make the error bound a little tighter by performing for example some grid search. In the same manner one can envision optimizing the parameter $\gamma$ of the estimator.  
\end{rem}
%%%%%%%%
\subsection{Laplacian kernel}
The Maximum Laplacian Correntropy estimator (MCE-L) corresponds to the case where the function $\ell$ in \eqref{eq:phi} is taken to be such that  $\ell(a)=|a|$. As a result, the function $\phi_\ell$  takes the form
\begin{equation}\label{eq:phi1}
	\phi_1(y,\hat{y})=\exp\left(-\gamma_1\left|y-\hat{y}\right|\right).
\end{equation}
\noindent It is straightforward to see that the properties P1-P4
are satisfied by $\ell$ with $\alpha_\ell=1$.  Theorem \ref{thm:main-theorem} can be specialized to this case as follows.  
%%%%%%% 
\begin{cor}[Laplacian kernel]\label{cor:Bound-exp-L1} $\: $\\
Let $I_\varepsilon^0$ be defined as in Theorem \ref{thm:main-theorem}.
Assume that the following condition is satisfied
\begin{equation}\label{eq:COND-MCE-L}
	\dfrac{1}{1+e^{-\gamma_1\varepsilon}}\rho_\alpha(X)+e^{-\gamma_1\varepsilon}\dfrac{|I_\varepsilon^0|}{N} >1
\end{equation}
for some  $\alpha \in \interval[open left]{0}{1}$. \\ 
Then for any $\theta^\star\in \argmax_{\theta\in \Re^n}\widehat{V}_{\phi_1}(y_t,\hat{y}_t(\theta))$ with $\phi_1$ defined as in \eqref{eq:phi1}, it holds that
\begin{equation}\label{eq:Bound-MCE-L}
	\left\|\theta^\star-\theta^o\right\|_2\leq \dfrac{1}{\gamma_1\alpha r_x}\ln\Big(\dfrac{1}{\mu(\gamma_1\varepsilon,\varepsilon,\alpha)}\Big)
\end{equation}
\end{cor}
%%%%%%%%%%%%%%%%%%%%

\subsection{Gaussian kernel}
The most used form of correntropy is the one based on the Gaussian kernel which, by omitting the normalizing factor, can be written in the form 
\begin{equation}\label{eq:phi2}
	\phi_2(y,\hat{y})=\exp(-\gamma_2 (y-\hat{y})^2)
\end{equation}
with $\gamma_2>0$. We will refer to the associated estimator as the maximum Gaussian correntropy estimator (MCE-G). Here, the function $\ell$ is defined by $\ell(a)=a^2$ and according to Lemma \ref{lem:examples}, it satisfies the properties P1-P4. In particular, P4 is satisfied with $\alpha_\ell=1/2$. Moreover $\ell$ is clearly monotonic on $\Re_+$.  As a consequence, we get a corollary of Theorem \ref{thm:main-theorem} as follows.  
\begin{cor}[Gaussian kernel]
Let $I_\varepsilon^0$ be defined as in Theorem \ref{thm:main-theorem}. % and $I_\varepsilon^c = \mathbb{I}\setminus I_\varepsilon^0$. 
Assume that the following condition is satisfied
\begin{equation}\label{eq:COND-MCE-G}
	\dfrac{1}{1+e^{-\gamma_2\varepsilon^2}}\rho_\alpha(X)+e^{-\gamma_2\varepsilon^2}\frac{|I_\varepsilon^0|}{N} >1
\end{equation}
for some  $\alpha \in \interval[open left]{0}{1}$.   \\
Then for any $\theta^\star\in \argmax_{\theta\in \Re^n}\widehat{V}_{\phi_2}(y_t,\hat{y}_t(\theta))$, it holds that
\begin{equation}\label{eq:Bound-MCE-G}
	\left\|\theta^\star-\theta^o\right\|_2\leq \dfrac{1}{\alpha r_x}\left[\dfrac{2}{\gamma_2}\ln\Big(\dfrac{1}{\mu(\gamma_2\varepsilon^2,\varepsilon,\alpha)}\Big)\right]^{1/2}.
\end{equation}
\end{cor}
%%%%%%%%%
%%%%%%%%%%%%%%%%%%%%
\subsection{A remark on the error-in-variables scenario}
We now consider the situation where only a noisy observation  $\bar{x}_t=x_t+w_t$ of the regressor vector $x_t$ in \eqref{eq:system} is available for prediction. This scenario is referred to as the robust error-in-variable (EIV) regression problem. 
Then the predictor output is given by 
$$
\hat{y}_t(\theta)=\bar{x}_t^\top \theta. 
$$
Indeed Theorem \ref{thm:main-theorem} remains valid for this case. To see this note that the system equation \eqref{eq:system} can be rewritten as 
$$y_t=\bar{x}_t^\top \theta^o+\bar{v}_t $$ 
where $\bar{v}_t=v_t-w_t^\top\theta^o$. Then clearly the theorem applies to the EIV scenario with $\left\{x_t\right\}$ and  $\left\{v_t\right\}$,  replaced respectively by $\left\{\bar{x}_t\right\}$ and $\left\{\bar{v}_t\right\}$. One limitation however in this case is that for a given $\varepsilon\geq 0$, the cardinality of the set $I_\varepsilon^0=\left\{t\in \mathbb{I}: \left|\bar{v}_t\right|\leq \varepsilon\right\}$ is likely to be much smaller than in the situation where the regressors are noise-free.   
%%%%%%%%%%%%%%%%%%%%%
\section{Numerical experiments}\label{sec:simulations}
The purpose of this section is to provide a numerical illustration of the richness measure \eqref{eq:Rho} and of the estimation error bound \eqref{eq:Error-Bound}. 
The system example considered for the experiment is of an FIR-type and is given by 
\begin{equation}
	%y_t = u_t+3u_{t-1}+ 2u_{t-2}+v_t
	y_t = 0.5u_t-u_{t-1}+ 0.2u_{t-2}+v_t
\end{equation}
which can be written in the form \eqref{eq:system} with $\theta^o=\bbm 0.5 & -1 & 0.2\eem^\top$ and $x_t = \bbm u_t & u_{t-1} & u_{t-2}\eem^\top$. 
For the data-generation experiment, assume that 
$\left\{u_t\right\}\stackrel{iid}{\sim} \mathcal{N}(0,1) $  i.e.,  $\left\{u_t\right\}$ is sampled independently and identically  from a zero-mean Gaussian distribution of unit variance. 
As for the noise signal $\left\{v_t\right\}$, it is defined as $v_t=e_t+f_t$ with $\left\{e_t\right\} \stackrel{iid}{\sim} \mathcal{U}(\interval{-\varepsilon}{\varepsilon})$ where $\mathcal{U}$ refers to the uniform distribution and $\left\{f_t\right\}$ is a sequence of sparse noise with only a few nonzero elements (which are otherwise not constrained in magnitude); the nonzero elements of $\left\{f_t\right\}$ are here sampled from $\mathcal{N}(50,10)$. 
%%%%

\subsection{Illustration of estimates}\label{subsec:estimates}
\begin{figure}[h!]
\centering
\psfrag{Laplacian}[][]{\tiny MCE-L}
\psfrag{Gaussian}[][]{\tiny MCE-G}
\psfrag{L1-norm}[][]{\tiny LAD}
\psfrag{Error}[][]{\small Error}
\psfrag{Epsilon}[][]{\small $\varepsilon$}
\includegraphics[scale=0.45]{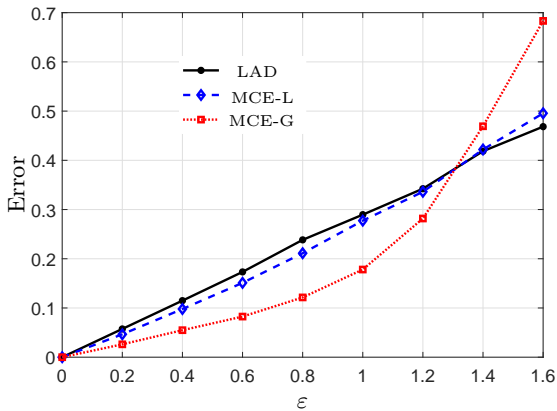} 
\caption{Comparison of the estimators MCE-L, MCE-G and LAD. Evolution of the (average)  estimation error
in function of the level $\varepsilon$ of noise. The range $\interval{0}{1.6}$ of $\varepsilon$ corresponds indeed to a range of about $\interval{2}{+\infty}$ for the signal-to-noise ratio in this experiment.  
The results are obtained from a Monte-Carlo simulation of size $1000$. For each experiment, the  proportion of outliers is maintained fixed and equal to $50\%$. Design parameters: $\gamma_1=0.5$ and $\gamma_2=0.25$. 
}
	\label{fig:comparison_mce_estimates}
\end{figure}

We generate $N=300$ data pairs $(x_t,y_t)$ and carry out a comparison between three estimators: on the one hand, the maximum Laplacian  correntropy  estimator (MCE-L) and the maximum Gaussian  correntropy  estimator (MCE-G) and on the other hand, the Least Absolute Deviation (LAD) estimator (which is also called $\ell_1$ estimator). Recall that MCE-G  and  MCE-L  involve non convex optimization. Here they are heuristically implemented  as a reweighted iterative least squares estimator and as a reweighted $\ell_1$ estimator respectively. 
The results are represented in Figure \ref{fig:comparison_mce_estimates} in term of average estimation error. What this suggests is that for fixed values of the design parameters $\gamma_1$ and $\gamma_2$ (see Eqs \eqref{eq:phi1} and \eqref{eq:phi2} for the roles of these parameters), LAD and MCE-L enjoy  a similar performance for small amount of noise. But as the noise level increases, LAD shows better  stability  capabilities than the MCE-L. Note that overall MCE-G tends to perform best in the setting of this experiment as long as the magnitude of the dense noise is reasonable (SNR larger than $2.8$ dB). {A possible justification for this  is that  squaring errors that contain outliers as in \eqref{eq:phi2} cancel out their influence more forcefully than just taking their absolute value as in \eqref{eq:phi1}.  } 
%%%

\subsection{Estimation of richness measure $\rho_\alpha(X)$}
\begin{figure}[h!]
\centering
\psfrag{alpha}[][]{$\alpha$}
\psfrag{rhomin}[][]{\scriptsize$\rho_\alpha(X)$}
\subfloat[$X\in \Re^{2\times 200}$]{\includegraphics[width=.4\textwidth,height=4cm]{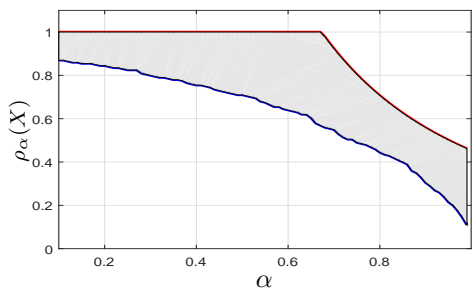}}\\
\subfloat[$X\in \Re^{3\times6000}$]{\includegraphics[width=.4\textwidth,height=4cm]{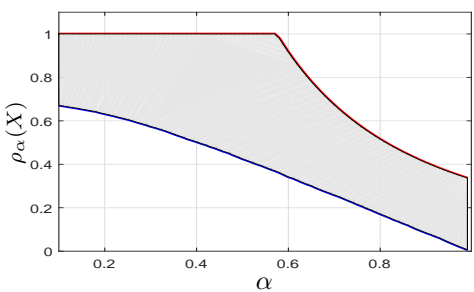}}
\caption{Estimates of $\rho_\alpha(X)$  as a function of $\alpha\in \interval{0}{1}$. 
The true values of $\rho_\alpha(X)$ lie in the region depicted in gray color. 
}
\label{fig:estimates-rho}
\end{figure}

We provide a graphical representation of how the informativity measure $\rho_\alpha(X)$ may, for a given data matrix $X\in \Re^{n\times N}$, evolve with respect to the dimensions $N/n$ of $X$ and the demanded degree $\alpha$ of richness  (See Figure \ref{fig:estimates-rho}). The estimated range for $\rho_\alpha(X)$ is based on Eq. \eqref{eq:Estimates-Rho}.   
Here $X$ is formed from an FIR-type of regressors with an input sampled from a zero-mean and unit variance Gaussian distribution.  
Our experiments in this specific study tend to suggest that $\rho_\alpha(X)$ is a non decreasing function of the ratio $N/n$ and a decreasing function of $\alpha$. Moreover, the estimated range (gray regions in Fig. \ref{fig:estimates-rho}) gets wider when $n$ is large. 
%%%%%%%%%%%%%
%%%%%%%%%%%%%%%%%

\subsection{Estimates of error  bounds}
\begin{figure}[h!]
\centering
\psfrag{Error}[][]{\small Bounds}
\psfrag{Epsilon}[][]{\small $\varepsilon$}
\psfrag{B1}{\tiny Bound MCE-L}
\psfrag{B2}{\tiny Bound MCE-G}
\includegraphics[scale=0.45]{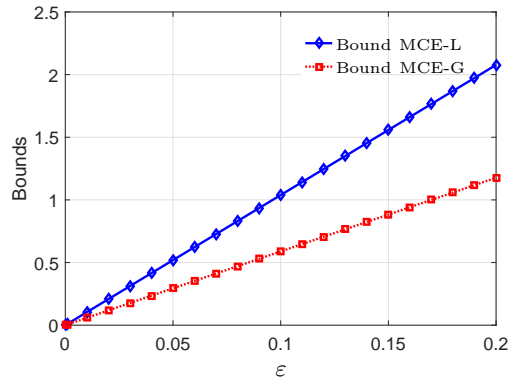}
\caption{Illustration of increasing rates of the error bounds. Results obtained by assuming that $\gamma\ell(\varepsilon)=0.2$, $|I_{\varepsilon}^0|/N=0.8$, $\rho_\alpha(X)=0.8$, $\alpha=0.6$ in both cases. These values are  chosen independently of a specific data matrix  $X$ but in such a way  that \eqref{eq:COND-MCE-L} and  \eqref{eq:COND-MCE-G} are satisfied.}
\label{fig:bounds}
\end{figure}

The goal here is twofold: (i) illustrate the variation of the estimation error bounds with respect to the magnitude of the dense noise    in the special cases \eqref{eq:Bound-MCE-L} and \eqref{eq:Bound-MCE-G}; (ii) assess how conservative the derived theoretical error bounds may be with respect to the empirical errors. \\
\paragraph{Increasing rates of the bounds}
 If for each level $\varepsilon$ of noise, we select the parameter $\gamma$ such that {the product $\gamma\ell(\varepsilon)$ is kept constant}, then the error bounds corresponding to both MCE-L and MCE-G have a linear rate of change with respect to $\varepsilon$ as depicted in Figure \ref{fig:bounds}. The increasing rate of the bound corresponding to MCE-L is larger than that of MCE-G for the current setting. Note that the computation of bounds made here is not connected to the experiment of Section \ref{subsec:estimates}.    
%%%%%%%%%%%%%*
%%%%%%%%%%%%%%%%%%%%%%%%%

\begin{figure}[h!]
\centering
%\psfrag{LogError}[][]{\small $\log_{10}(\mbox{\tiny Error})$}
\psfrag{LogError}[][]{\scriptsize $\log_{10}(\left\|\theta^\star-\theta^o\right\|_2)$}
\psfrag{N}[][]{\small $N$}
\psfrag{Laplacian}{\tiny MCE-L}
\psfrag{Gaussian}{\tiny MCE-G}
\psfrag{B1}{\tiny Bound MCE-L}
\psfrag{B2}{\tiny Bound MCE-G}
\includegraphics[scale=0.45]{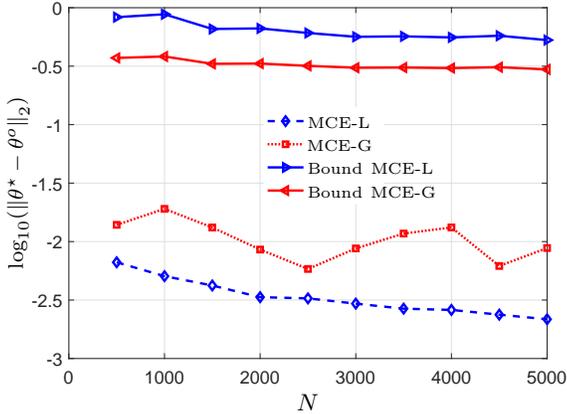} 
\caption{Comparison of theoretical bounds and empirical errors $\left\|\theta^\star-\theta^o\right\|_2$. 
Results obtained with an  estimate of $\rho_\alpha(X)$ for $\alpha=0.6$,  $\varepsilon=0.05$ (i.e., SNR $=25$ dB),  $\gamma\ell(\varepsilon)=0.2$, $|I_{\varepsilon}^0|/N=0.9$. The results are obtained from a Monte-Carlo simulation of size $1000$. 
}
\label{fig:conservativeness}
\end{figure}

\paragraph{Comparing theoretical bounds and empirical errors}
It might be instructive to see how far away the theoretical error bounds may be from the empirical values. To study  this aspect, let us consider a numerical experiment with a similar data-generating process as described in the beginning of Section \ref{sec:simulations}. The dense noise level is set to $\varepsilon=0.05$ which gives an SNR of about $25$ dB and the proportion of outliers is set to  $10\%$ (which is small enough to enforce condition \eqref{eq:Assump-INEQ}). 
One difficulty in evaluating the theoretical bounds is that this requires evaluating $\rho_\alpha(X)$ which, as already discussed in Section  \ref{subsec:Informativity}, is a hard problem. Hence,  $\rho_\alpha(X)$ is replaced here  with the mean value of the lower and upper estimates displayed in \eqref{eq:Estimates-Rho}. 
We then let the number $N$ of data  vary from $500$ to $5000$ and plot the empirical errors along with the bounds from \eqref{eq:Bound-MCE-L} and \eqref{eq:Bound-MCE-G} in Figure \ref{fig:conservativeness}.    
%%%%
%%%%%%%

It is fair to observe that the theoretical bounds are conservative in the sense that they are generally higher than the true empirical errors.  Here the ratio between the bounds and the true errors is about $30$.   Conservativeness is indeed a common feature for these types of results due to the various inequalities employed for the derivation. Nevertheless, the main interest of Theorem \ref{thm:main-theorem} is that it provides a sufficient condition for the robustness of the maximum correntropy estimator, a condition that depends explicitly on the degree of informativity of the regression data and on the proportion of outliers. Moreover, by expressing error  bounds 
which involve explicitly the design parameters, the theorem gives insights into how to tune those parameters with the aim to improve estimation performance. \\
A further remark one can make is that the general formula for the error bound in \eqref{eq:Error-Bound}  has a kind of universal feature in the following sense: since the bound does not involve the magnitude of the true $\theta^o$ (for an FIR-type system for example), it is  in principle valid regardless of $\theta^o$. Hence the relative error will be as smaller as the {norm} of the to-be-estimated parameter vector $\theta^o$ is larger.

%%%%%%%%%%%%%%%%%%%%%%%%%%%%%%%%%%%%%%%%%%%%%%%%%%%%%%	
\section{Conclusion}\label{sec:conclusion}
In this paper we have proposed an analysis of the robustness properties of a correntropy maximization framework for regression problems. 
The class of estimators considered is quite general and include the Gaussian and Laplacian kernels as special cases. The contribution of the work consists in  (i) deriving an appropriate notion of richness for the regression data; (ii) proving stability of the considered class of estimators under the derived richness condition when the data are subject to dense and sparse noise (outliers).  Our main result	states that if the regression data are rich enough and if the number of outliers is small in some sense, then the parametric estimation error is bounded. The results come with  explicit bounds which, in default of being exactly computable, can be estimated with computable estimates. 
	
	%%%%%%
	\begin{ack}
The author is grateful to the Associate Editor and the anonymous
reviewers for constructive feedback.
\end{ack}

\bibliographystyle{plain}
%\bibliography{references}
%%%%

\end{document}